\documentclass[preprint]{aastex}
\usepackage{epsfig}
 
\newcommand {\be} {\begin{equation}}
\newcommand {\ee} {\end{equation}}

\begin{document}
 
\title{A Model for the Optical/Infrared Emission from Magnetars}
 
 \author{Feryal \"Ozel\altaffilmark{1}}
\affil{University of Arizona, Departments of Physics and Astronomy\\ 
1118 E. 4th St., Tucson, AZ 85721; fozel@physics.arizona.edu }
\altaffiltext{1}{Hubble Fellow}
 
\begin{abstract}

A number of Anomalous X-ray Pulsars (AXPs) have recently been detected
in the optical/IR wavelengths. We use their inferred brightness to
place general constraints on any model for this emission within the
magnetar framework. We find that neutron-star surface emission cannot
account for the observations and that the emission must be
magnetospheric in origin. We propose a model for the optical/IR
emission in which a distribution of energetic electrons in the
neutron-star magnetosphere emits synchrotron radiation. This model can
naturally reproduce the observed brightness and the rising spectra of
AXPs as well as the observed pulsations at the stellar spin frequency
and the correlation of the IR flux with their bursting activity.

\end{abstract}
 
\section{Introduction}

Magnetars are a class of neutron stars powered by the decay of their
ultrastrong magnetic fields (Duncan \& Thompson 1992). If they are
isolated, magnetars are believed to appear as persistent pulsars in
the X-rays, with occasional episodes of bursting activity in the hard
X-rays/soft $\gamma$-rays. Anomalous X-ray Pulsars (AXPs) and Soft
Gamma-ray Repeaters (SGRs) are thought to be the observational
manifestations of magnetars: their steady spin-down, quasi-thermal
X-ray spectra, bursts, and the absence of binary companions all lend
support to this identification (see, e.g., Kouveliotou et al.\ 1998;
Mereghetti, Israel, \& Stella 1998; Woods et al.\ 1999; \"Ozel,
Psaltis, \& Kaspi 2001; Gavriil \& Kaspi 2002; Gavriil, Kaspi, \&
Woods 2003). The absence, until recently, of detectable emission from
AXPs and SGRs in longer wavelengths constrained all such studies to
the X-rays and $\gamma$-rays.

In the last few years, faint counterparts of four AXPs have been
detected in IR and optical wavelengths (Hulleman, van Kerkwijk, \&
Kulkarni 2000, 2004; Hulleman et al.\ 2001; Israel et al.\ 2002, 2003;
Wang \& Chakrabarty 2002) but no counterparts have yet been observed
for SGRs, due at least in part to the high extinction along the line
of sight to these sources (Corbel et al.\ 1997, 1999; Vrba et al.\
2000; Kaplan et al.\ 2001, 2002; Eikenberry et al.\ 2001; Wachter et
al.\ 2004). The AXP counterparts were found to be variable, with
fluxes possibly correlated with the bursting activity (Kaspi et al.\
2003), had rising spectra in $\nu F_\nu$ (Hulleman et al.\ 2004; see
Fig.~1 below), and showed pulsations at the stellar spin frequency
(Kern \& Martin 2002). These long-wavelength detections have been used
to argue strongly against the presence of accretion disks and
companion stars around AXPs (Hulleman et al.\ 2000; Perna, Hernquist,
\& Narayan 2000). However, no predictions have been made so far for
the expected long-wavelength emission from a magnetar, which can be
tested against observations (see Eichler, Gedalin, \& Lyubarsky 2002
for a preliminary discussion).

In this {\it Letter}, we discuss the strong constraints imposed by the
observed IR and optical brightness of AXPs on any magnetar emission
model. In particular, we argue that the long-wavelength emission
cannot originate from the surface of the neutron star. We suggest that
synchrotron emission from energetic particles in the magnetosphere at
$\gtrsim 50$ stellar radii is responsible for this emission.

\section{Constraints on Emission Models}

The brightness and limits inferred from the optical and IR
observations of AXPs and SGRs can be used to place stringent
constraints on the mechanisms that can give rise to this emission
within the magnetar model. The most robust constraint comes from
imposing the thermodynamic limit on the surface emission from a
neutron star.

If the source of energy that powers the optical/IR emission is in the
interior of the neutron star, the emitted radiation will be
reprocessed in the thermalized surface layers of the star. For such
processes, the maximum flux that can emerge from the star is equal to
the flux of blackbody radiation at the temperature $T_{\rm BB}$ at
optical depth unity. For a source at a distance of 5~kpc observed at a
frequency of $10^{15}$~Hz, the blackbody limit is
\begin{equation}
\nu F_{\nu,{\rm BB}} = 5\times 10^{-18} 
\left(\frac{R_{\rm NS}}{10^6~{\rm cm}}\right)^2
\left(\frac{D}{5~{\rm kpc}}\right)^{-2}
\left(\frac{T_{\rm BB}}{0.1~{\rm keV}}\right)
\left(\frac{\nu}{10^{15}~{\rm Hz}}\right)^3
{\rm erg~s}^{-1}~{\rm cm}^{-2}, 
\end{equation}
where $R_{\rm NS}$ is the radius of the neutron star. Models of
magnetar atmospheres indicate that the temperature at the
thermalization depth of optical/IR photons is $T_{\rm BB} \approx
0.1$~keV in an atmosphere of $T_{\rm eff} \approx 0.5$~keV, which
reproduces the X-ray spectra of AXPs (\"Ozel 2001, 2003).

The constraint imposed on the emission models by equation~(1) is
illustrated in Figure~1 for the case of AXP 4U~0142$+$61, which has
been detected over the widest range of optical/IR wavelengths. This
source has a period of $P=8.69$~s, a period derivative of $\dot{P} =
2\times10^{-12}$~s~s$^{-1}$, and an uncertain distance of $D>1$~kpc or
$D>2.7$~kpc (see \"Ozel et al.\ 2001 and references therein). For this
discussion, we set $D=2$~kpc.

\begin{figure}[t]
\centerline{ \psfig{file=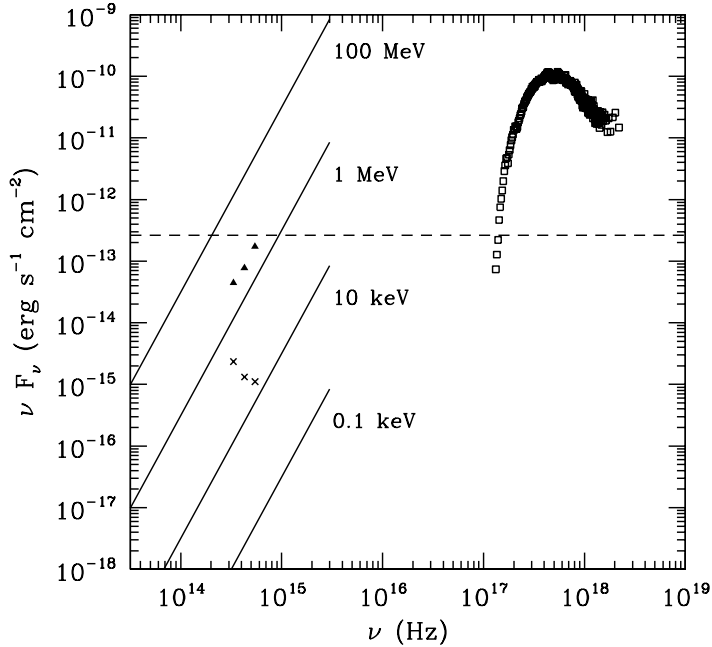,width=11truecm} }
\figcaption[]{ The solid lines show the blackbody limit for different
temperatures on the neutron star surface for a source at 2~kpc. The
dashed line shows the maximum rotation-powered flux for the timing
parameters of 4U~0142$+$61. The crosses and triangles show the
absorbed and unabsorbed fluxes (Hulleman et al.\ 2000), respectively,
inferred for 4U~0142$+$61, while the squares show the absorbed X-ray
flux. 
\label{Fig:bb}}
\end{figure}

As Figure~1 shows, the maximum possible flux from the surface of the
neutron star 4U~0142$+$61 is smaller by orders of magnitude than the
observed optical flux for any reasonable value of the surface
temperature. Indeed, for the blackbody limit to be comparable to the
optical flux, the temperature of the emitting layers needs to exceed
1~MeV. At this temperature, the cooling of the neutron star would
proceed via neutrino and not photon emission. Moreover, it would yield
an X-ray spectrum inconsistent with the observations. These strongly
argue against a magnetar surface emission model for the optical/IR
emission of AXPs.

An upper limit on the size of the emitting region can be obtained by
considering the large amplitude of pulsations at the stellar spin
frequency observed at optical wavelengths for 4U~0142$+$61 (Kern \&
Martin 2002). For coherent pulsations to be produced at $P=8.69$~s,
the size of the emitting region has to be smaller than the light
travel distance $R < cP \simeq 10^5 R_{\rm NS}$, which also coincides
with the size of the light cylinder. Therefore, the optical/IR
emission of AXPs must be magnetospheric in origin. Note that these
limits apply in either the case of emission from or reprocessing of
stellar radiation by the magnetosphere.

The source of energy that powers the optical/IR emission is more
difficult to constrain. However, contrary to the case of the
high-energy emission which requires a source of magnetic energy, it is
significant that rotational energy alone is sufficient to account for
the observed long-wavelength brightness of AXPs. This is illustrated
in Figure~1 for the case of 4U~0142$+$61, where the horizontal line
corresponds to the maximum flux that can be observed at Earth from
processes powered by rotational energy losses, i.e.,
\begin{eqnarray}
\nu F_{\nu,{\rm rot}} &=&\frac{\dot{E}_{\rm rot}}{4 \pi D^2}\nonumber \\
&=& 2.7\times 10^{-13} 
\left(\frac{I}{10^{45}~{\rm g~cm}^2}\right)
\left(\frac{D}{2~{\rm kpc}}\right)^{-2} \nonumber \\
&& \left(\frac{P}{8.69~{\rm s}}\right)^{-3}
\left(\frac{\dot P}{2 \times 10^{-12}~{\rm s~s}^{-1}}\right)
{\rm erg~s}^{-1}~{\rm cm}^{-2}. 
\end{eqnarray}
In this equation, $\dot{E}_{\rm rot}$ is the rotational luminosity and
$I$ is the moment of inertia of the neutron star. Further observations
of AXPs in UV wavelengths may be able to distinguish between the
magnetic and rotational energy as the source of the long-wavelength
emission.

\section{A Magnetospheric Model for Long-Wavelength Emission}

In the previous section, we argued that the optical/IR emission from
AXPs should originate within their magnetospheres. In this section, we
show that synchrotron emission from a Goldreich-Julian density of
electrons in a dipole magnetic field geometry at $\gtrsim 50$~neutron
star radii can reproduce the observed long-wavelength properties of
AXPs.

In the magnetosphere of a rotating neutron star, the quasi steady-state
distribution of charges is given by (Goldreich \& Julian 1969)
\begin{equation}
N(r) = \frac{B(r)}{e c P}, 
\end{equation}
where $B(r)$ is the local magnetic field at radius $r$, $e$ is the
electric charge, and $c$ is the speed of light. The acceleration
mechanism and the resulting energy distribution of these charges is
largely unknown. In this section, we will assume a mono-energetic
distribution of electrons characterized by a Lorentz factor $\gamma$.
We will also assume a magnetic field strength with a dipolar radial 
profile
\begin{equation}
B(r)=B_{\rm NS}\left(\frac{R_{\rm NS}}{r}\right)^3
\end{equation}
where $B_{\rm NS}$ is the magnetic field strength at the stellar
surface.

This distribution of energetic charges in the stellar magnetic field 
emits synchrotron radiation with an integrated power 
\begin{equation}
P_{\rm tot} = \frac{4}{3} \sigma_{\rm T} c N(r) \beta^2 \gamma^2 
\frac{B^2}{8\pi},
\end{equation}
where $\sigma_{\rm T}$ is the Thompson cross-section and $\beta$ is
the velocity of the electrons in units of $c$. In this calculation, 
we assume that, at each radius, all the power is emitted at the 
fundamental synchrotron frequency
\begin{equation}
\nu_c(r) = \frac{eB(r)}{\gamma m_e c} = 
\frac{2 \times 10^{21}}{\gamma} 
\left(\frac {B}{10^{14}~{\rm G}}\right)
\left(\frac {R_{\rm NS}}{r}\right)^3~{\rm Hz}, 
\end{equation}
where $m_e$ is the mass of the electron. Note that the classical
expression for the synchrotron emissivity is sufficient for the
magnetic field strengths that generate the optical/IR frequencies.

Under these assumptions, the photon-frequency dependent luminosity 
can be calculated as
\begin{equation}
L_\nu = \int_{R_{\rm NS}}^\infty P_{\rm tot}(r) \delta[\nu - \nu_c(r)]
4 \pi r^2 dr, 
\end{equation}
which yields
\begin{equation}
L_\nu = 1.1 \times 10^{16} \beta^2 \gamma^4 
\left(\frac{P}{5~{\rm s}}\right)^{-1}
\left(\frac{B_{\rm NS}}{10^{14}~{\rm G}}\right)
\left(\frac{\nu}{10^{15}~{\rm Hz}}\right)
{\rm erg~s}^{-1}~{\rm Hz}^{-1}. 
\end{equation}
This luminosity corresponds to a flux at Earth of 
\begin{equation}
\nu F_\nu = 3.5 \times 10^{-15} \beta^2 \gamma^4 
\left(\frac{P}{5~{\rm s}}\right)^{-1}
\left(\frac{B_{\rm NS}}{10^{14}~{\rm G}}\right)
\left(\frac{D}{5~{\rm kpc}}\right)^{-2}
\left(\frac{\nu}{10^{15}~{\rm Hz}}\right)^2
{\rm erg~s}^{-1}~{\rm cm}^{-2}. 
\end{equation}

\begin{figure}[t]
\centerline{ \psfig{file=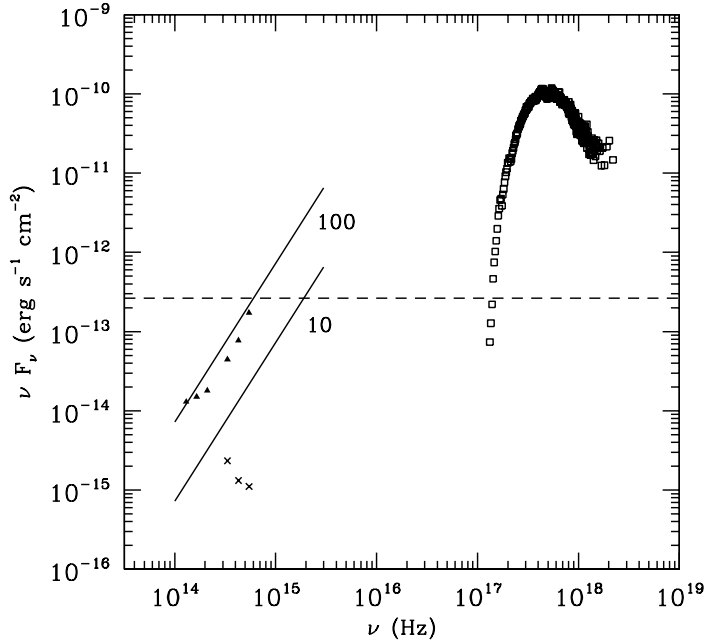,width=11truecm} }
\figcaption[]{The Optical/IR spectrum for magnetospheric synchrotron
emission from a magnetar. Labels 10 and 100 refer to the values for
the parameter $\beta^2 \gamma^4 (B/10^{14}~{\rm G})$. The data points
and the dashed line are as in Figure~1.
\label{Fig:ir}}
\end{figure}

The predictions of this model for the case of 4U~0142$+$61 are shown
in Figure~2. The optical spectrum of this source, which approximately
follows a $\nu^2$ dependence, can be fit very well with a
mono-energetic distribution of electrons characterized by 
\begin{equation}
\gamma \simeq 3 \left(\frac{B_{\rm NS}}{10^{14}~{\rm
G}}\right)^{-1/4}.
\end{equation}
We discuss the effects of relaxing some of the approximations made in
this section and its implications for the IR spectra of AXPs in the
next section.

\section{Discussion}

In this {\em Letter}, we described a model of the optical/IR emission
of AXPs in which a distribution of energetic electrons in the neutron
star magnetosphere emits synchrotron radiation. This model can
naturally reproduce the observed brightness and the rising spectra of
AXPs.

In the calculations reported above, we have made some simplifying
assumptions that may affect the quantitative predictions of the
model. First, we did not take into account the angular dependence of
the magnetic field and of the emitted radiation. Relaxing this
assumption will result in a brightness and spectrum that depends on
the relative orientation of the magnetic axis and the
observer. Moreover, it will naturally give rise to pulsations of the
long-wavelength emission at the stellar spin frequency. Second, we
have assumed a mono-energetic distribution of electrons throughout the
magnetosphere. Allowing for a radial profile of the electron Lorentz
factor $\gamma$ will affect the slope of the optical/IR spectrum and
may give rise to spectral breaks at frequencies characteristic of the
acceleration and cooling mechanisms in the magnetosphere.

The source of energy and the mechanism of particle acceleration in the
magnetosphere of a magnetar are open questions. The long-wavelength
observations of AXPs may provide within this model clues towards
distinguishing between various possibilities. If the source of this
radiation is dissipation of the rotational energy of the neutron star,
then equation~(2) provides an upper limit on the flux of
magnetospheric synchrotron photons (see the dashed line in Fig.~2).
This, in return, sets an upper limit on the frequency of synchrotron
radiation at $\simeq 10^{15}$~Hz for the parameters of 4U~0142$+$61.
Combining this limit with equation~(6) and the best-fit value of the
parameter $B_{\rm NS} \gamma^4$ (see Fig.~2) results in a lower limit
on the emission radius of $r/R_{\rm NS} \gtrsim 80 (B_{\rm
NS}/10^{14}$~G$)^{5/12}$. The observations of 4U~0142$+$61, in this
framework, require that either the particles do not radiate at smaller
radii or that particle acceleration takes place at large distances
from the neutron star surface, possibly due to the formation of bound
positronium states (Leinson \& P\'erez 2000) at the stronger magnetic
fields close to the star.

The inferred IR spectrum of 4U~0142$+$61 is flatter than its optical
spectrum (Hulleman et al.\ 2004). This can be achieved for a radial
distribution of Lorentz factors that rises locally with increasing
radius. Such an energy distribution would strongly suggest that the
particles are accelerated at $\gtrsim 100 R_{\rm NS}$ and lose energy
via synchrotron radiation as they follow field lines back toward the
star. If the low-frequency radiation is powered by magnetic energy,
such a configuration may naturally arise from the dissipation of
Alfv\'en waves in the magnetospheres of magnetars (Thompson \& Duncan
1996). Because of the much larger reservoir of magnetic energy, the
optical spectrum may extend to higher frequencies than in the case of
rotation power. Observations of AXPs and SGRs in UV wavelengths will
be important in distinguishing between different mechanisms. Note also
that the optical and IR observations of 4U~0142$+$61 were not carried
out simultaneously and may point to a change in the magnetic field
configuration in time. Simultaneous and/or repeat observations may
also provide additional clues to the nature of the emission.

This magnetospheric model can also account for a number of other
observed properties of the optical/IR emission of AXPs. As mentioned
above, the magnetic field geometry and the angular dependence of
synchrotron emission gives rise to pulsations at the stellar spin
frequency as observed (Kern \& Martin 2002). The optical/IR flux is
also likely to be polarized due to the strong beaming of synchrotron
emission but a quantitative prediction for the degree of polarization
requires including in the calculations the angle dependences discussed
above. Observations that search for polarization of the optical
emission may help further constrain the magnetospheric model. In
addition, the optical/IR emission will naturally respond to the
changes in the magnetic field strength, configuration, or spin-down
rate following an SGR-like burst. The timescale for this flux
enhancement will be dictated by the rearrangement of magnetic field
lines which are anchored in the neutron star crust and thus will
proceed at a slower pace than any characteristic timescale in the
magnetosphere. Such correlated changes have been observed following
the burst of 1E~2259$+$586 (Kaspi et al.\ 2003).

\begin{figure}[t]
\centerline{ \psfig{file=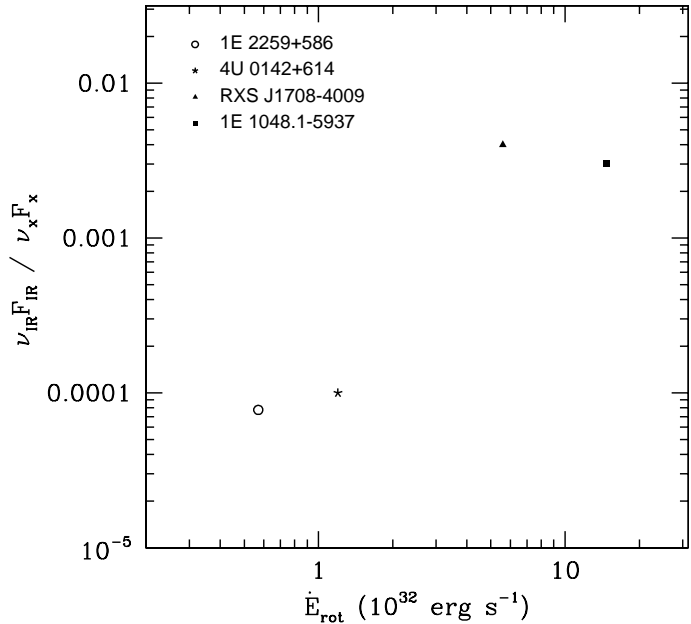,width=11truecm} } \figcaption[]{The
ratio of the unabsorbed IR flux in the K band to the X-ray flux at $5
\times 10^{17}$~Hz as a function of the rotational luminosity for four
persistent AXPs. Normalizing to the X-ray flux minimizes the
differences in distance uncertainties and absorption column between
the sources. The positive correlation supports the idea of
magnetospheric origin for the long-wavelength emission.
\label{Fig:rot}}
\end{figure}

Finally, because of its proposed magnetospheric origin, the low-energy
emission of AXPs is expected to correlate with the rate of rotational
or magnetic energy losses. The absolute luminosities of AXPs in the IR
are very hard to infer observationally because of the unknown
distances and the effects of the large interstellar absorption to
these sources. To minimize these uncertainties, we have normalized the
unabsorbed IR flux in the K band to the unabsorbed persistent X-ray
flux at $5 \times 10^{17}$~Hz. The intrinsic X-ray luminosities of
AXPs are expected to be similar between the sources and not to depend
strongly on the magnetic field strength. Even though such a luminosity
clustering is difficult to infer from the data, their quasi-thermal
spectra with similar color temperatures and the fact that the emitting
areas should all be comparable to the surface area of a neutron star
point to such a theoretical expectation. The inferred flux ratios are
plotted against the observed rate of rotational energy loss in
Figure~3 (see Israel et al.\ 2003 and references therein for the
data). The overall correlation provides additional support to the
model of magnetospheric optical/IR emission of AXPs discussed here.

\acknowledgements I thank Chryssa Kouveliotou, Peter Woods, and
Sandeep Patel for stimulating discussions and their hospitality during
my visit to MSFC where this work was initiated, as well as for their
detailed comments on the manuscript. I thank Sandeep Patel also for
his help with the AXP data files. I am grateful to Dimitrios Psaltis
for his help, scientific and otherwise, while completing this work. I
acknowledge support by NASA through Hubble Fellowship grant HF-01156
from the Space Telescope Science Institute, which is operated by the
Association of Universities for Research in Astronomy, Inc., under
NASA contract NAS 5-26555.

\end{document}